\documentclass[preprint,double-spaced]{revtex4}
\usepackage{wrapfig}
\usepackage{graphicx}
\usepackage{amsmath,bbm}
\usepackage{amssymb,bm}

\makeatletter

\newcommand{\Rmnum}[1]{\expandafter\@slowromancap\romannumeral #1@}
\makeatother

\begin{document}

\title{The effect of flow on Hadronic Spectra in an Excluded-Volume Model}
\author{S.~K.~Tiwari\footnote{corresponding author: $sktiwari4bhu@gmail.com$}}
\author{P.~K.~Srivastava}
\author{C.~P.~Singh}

\affiliation{Department of Physics, Banaras Hindu University, 
Varanasi 221005, INDIA}

\begin{abstract}
\noindent
Recently, we proposed a thermodynamically consistent excluded-volume model for the HG fireball and we noticed that our model gives a suitable description for various properties of multiparticle production and their ratios in the entire range of temperatures and baryon densities. The aim in this paper is to obtain the variations of freeze-out volume in a slice of unit rapidity i.e. $dV/dy$ as well as total volume of the fireball with respect to center-of-mass energy $(\sqrt{s_{NN}})$ and confront our model calculations with the corresponding thermal freeze-out volume obtained from the Hanbury-Brown-Twiss (HBT) pion interferometry method. We also test the validity of our model in extracting the total multiplicities as well as the central rapidity densities of various hadrons and comparing them with the recent results. We further calculate the rapidity as well as transverse momentum spectra of various particles produced in different heavy-ion collider experiments in order to examine the role of flow by matching our predictions suitably with the available experimental results. Finally, we extend our analysis for the production of light nuclei, hypernuclei and their antinuclei over a broad energy range from Alternating Gradient Synchrotron (AGS) to Large Hadron Collider (LHC) energies.


\end{abstract}

\maketitle 

\section{Introduction}
\noindent
The ultimate goal of ultra-relativistic heavy-ion collisions is to produce a highly excited and dense matter which possibly involves a phase transition from a hot, dense hadron gas (HG) to a deconfined quark matter called as quark-gluon plasma (QGP) \cite{Singh:1993,Satz:2000,Shuryak:1980}. In this state the degrees of freedom are those of quarks and gluons only. One hopes that by colliding heavy nuclei, one can create a fireball with an extremely large energy density extending over a sufficiently large space-time volume so that an equilibrated quark-gluon plasma may be formed. However, experimental and theoretical investigations made so far reveal that it is indeed difficult to get an unambiguous evidence for QGP formation. It is very important to understand properly the dynamics of the collisions in order to suggest any unique signal for QGP. Such information can be obtained by analyzing the properties of various particles which are emitted from various stages of the collisions. Hadrons are produced at the end of the hot and dense QGP phase, but they subsequently scatter in the confined hadronic phase prior to decoupling (or "freeze-out") from the collision system and finally a collective evolution of the hot and dense matter occurs in the form of transverse, radial or elliptic flow which are instrumental in shaping the important features of particle spectra. The global properties and dynamics of freeze-out can be at best studied via hadronic observables such as hadron yields, ratios, rapidity distributions and transverse mass spectra \cite{Letessier:2004}.

Recently various types of statistical thermal models have been used for the description of hot, dense HG. But excluded-volume corrected models \cite{Cleymans:1986} have attracted more attention due to their capability of describing lattice QCD data for various physical quantities \cite{Andronic:2012}. Recently we have proposed a new excluded-volume model \cite{Tiwari:2012} and studied different thermodynamical properties such as number density, energy density, pressure etc. of HG besides deriving some freeze-out conditions for the description of the decoupling stage from the fireball. It is indeed surprising that the predictions of our geometrical model regarding the detailed features of various hadron ratios and their variations with respect to the center-of-mass energy ($\sqrt{s_{NN}}$) are successfully tested with the available experimental data upto Relativistic Heavy-Ion Collider (RHIC) energies. We have also predicted the hadron yields which we expect at the Large Hadron Collider (LHC) energies. We further notice that the model respects causality \cite{Tiwari:2012} and the values of transport coefficients such as shear viscosity-to-entropy ratio ($\eta/s$) and the speed of sound as obtained from our model match with the predictions of other HG models. Our aim in this paper is to extend our model \cite{Tiwari:2012} to calculate rapidity as well as transverse mass spectra of various hadrons. In heavy-ion collisions, rapidity densities of produced particles are strongly related to the energy density and/or entropy density created in the collisions \cite{Busza:1988}. Further, the dependence of transverse mass spectra of hadrons on $\sqrt{s_{NN}}$ can also yield insight into the evolution of a radial flow present in the dense fluid formed in the collision \cite{Blume:2011}. Earlier different types of approaches have been used for studying the rapidity and transverse mass spectra of hadrons \cite{Landau:1953,Schnedermann:1993,Braun:1996,Schnedermann:1992,Feng:2011,Uddin:2012,Hirano:2002,Manninen:2011,Bass:1998,Mayer:1997,Becattini:2007,Biedron:2007,Cleymans:2008,Broniowski:2001,Adler:2001,Adcox:2004,Bozek:2008}. Hadronic spectra from purely thermal models usually reveal an isotropic distribution of particles \cite{Schnedermann:1993} and hence the rapidity spectra obtained with the thermal models do not reproduce the features of the experimental data satisfactorily. Similarly, the transverse mass spectra from the thermal models reveal a more steeper curve than that observed experimentally. Comparisons thus illustrate that the fireball formed in heavy-ion collisions does not expand isotropically in nature and there is a prominent input of collective flow in the longitudinal and transverse directions resulting in an anisotropy in the rapidity and transverse mass spectra of the hadrons after the freeze-out. These results suggest the necessity for the inclusion of a flow factor in the excluded-volume model in order to reproduce  the rapidity as well as transverse mass spectra of various hadrons \cite{Manninen:2011}. In this paper, we attempt to calculate the rapidity density of various particles at midrapidity by using our excluded-volume model of HG without any flow effect and comparisons yield a good agreement between our results and the experimental data. However, when we calculate the complete rapidity distributions of the particles, we find that the distribution always has a narrow shape in comparison to the experimental data and hence our thermal model alone is incapable to describe the experimental data in forward and backward rapidity regions. In a similar way, the transverse mass spectra of hadrons as obtained in our model again does not properly match with the experimental data. However, after incorporation of an additional flow-effect as suggested in Ref. \cite{Schnedermann:1993}, we notice that our model suitably describes the data.

The plan of the paper runs as follows : the next section deals with the formulation of our model to be used in the study of the rapidity and transverse mass spectra of hadrons using purely thermal source. In section III, we modify the formula for the rapidity distributions by incorporating a flow velocity in the longitudinal direction and similarly the formula for the transverse mass spectra is also modified by incorporating a collective flow in the longitudinal as well as in the transverse direction. In section IV, we compare the experimental data with our predictions regarding rapidity density, transverse mass spectra and transverse momentum spectra of hadrons at RHIC $(200\;GeV)$ and LHC energy $(2.76\;TeV)$. More importantly we have also deduced total mean multiplicities, mid-rapidity yields of various hadrons and their rapidity densities and variations with $\sqrt{s_{NN}}$ yield a good fit to the experimental data. Finally the freeze-out volume and/or $dV/dy$ are determined in order to ascertain whether the emissions of all hadrons occur from the same hypersurface of the fireball. In section V, we also analyze the experimental data for the production of light nuclei, hypernuclei and their antiparticles over a broad energy range going from AGS to RHIC energies and we again compare them with our predictions. This analysis thus fully demonstrates the validity of our EOS in describing various features of a hot, dense HG. Finally, in the last section we succinctly give the conclusions and summary.

\section{Hadronic Spectra with the thermal source}
\noindent 

\subsection{Rapidity Distributions}

To study the rapidity distributions and transverse mass spectra of the hadrons, we extend excluded-volume model as proposed in Ref. \cite{Tiwari:2012}, which was found to describe the hadron ratios and yields at various $\sqrt{s_{NN}}$ from lower AGS energies upto RHIC energies with remarkable success. We have incorporated the excluded-volume correction directly in the grand canonical partition function of HG in a thermodynamically consistent manner. We obtain the number density $n_i^{ex}$ for ith species of baryons after excluded-volume correction as follows \cite{Tiwari:2012}:

\begin{equation}
n_i^{ex} = (1-R)I_i\lambda_i-I_i\lambda_i^2\frac{\partial{R}}{\partial{\lambda_i}}+\lambda_i^2(1-R)I_i^{'},
\end{equation}
where $\displaystyle R=\sum_in_i^{ex}V_i^0$ is the fractional occupied volume by the baryons \cite{Tiwari:2012}. $\displaystyle V_i^0= 4\pi\;r'^3/3$ is the eigen-volume of each baryon having a hard-core radius $r'$ and $\lambda_i$ is the fugacity of ith baryon. Further, $I_i$ is the integral of the baryon distribution function over the momentum space \cite{Tiwari:2012}. 

We can rewrite Eq. (1) in the following manner :

\begin{eqnarray}
\begin{aligned} 
n_i^{ex} 
=\frac{g_i\lambda_i}{(2\pi)^3}\;\Big[\Big((1-R)-\lambda_i\frac{\partial{R}}{\partial{\lambda_i}}\Big)\;\int_0^\infty \frac{d^3k}{\displaystyle \Big[exp\left(\frac{E_i}{T}\right)+\lambda_i \Big]}
\\
-\lambda_i(1-R)\;\int_0^\infty \frac{d^3k}{\displaystyle \Big[exp\left(\frac{E_i}{T}\right)+\lambda_i\Big]^2}\Big].
\end{aligned} 
\end{eqnarray}

This reveals that the invariant distributions are \cite{Schnedermann:1993,Braun:1996} :

\begin{eqnarray}
\begin{aligned} 
E_i\;\frac{d^3N_i}{dk^3}=\frac{g_iV\lambda_i}{(2\pi)^3}\;\Big[\Big((1-R)-\lambda_i\frac{\partial{R}}{\partial{\lambda_i}}\Big)\; \frac{E_i}{\displaystyle \Big[exp\left(\frac{E_i}{T}\right)+\lambda_i\Big]}\\
-\lambda_i(1-R)\;\frac{E_i}{\displaystyle \Big[exp\left(\frac{E_i}{T}\right)+\lambda_i\Big]^2}\Big].
\end{aligned}
\end{eqnarray}

In case we use Boltzmann's approximation, our Eq. (3) differs from the one used in the paper of Schnedermann $et\;al.$ \cite{Schnedermann:1993} by the presence of a prefactor $\displaystyle\Big[(1-R)-\lambda_i\frac{\partial{R}}{\partial{\lambda_i}}\Big]$. However, all these quantities are determined precisely at the chemical freeze-out in our model and hence quantitatively we do not require any normalising factor as is required in Ref. \cite{Schnedermann:1993}.

Using :

\begin{eqnarray}
E_i\;\frac{d^3N_i}{dk^3}&=&\frac{dN_i}{dy\;m_T\;dm_T\;d\phi},
\end{eqnarray}

we get :

\begin{eqnarray}
\begin{aligned} 
\frac{dN_i}{dy\;m_T\;dm_T\;d\phi}=\frac{g_iV\lambda_i}{(2\pi)^3}\;\Big[\Big((1-R)-\lambda_i\frac{\partial{R}}{\partial{\lambda_i}}\Big)\; \frac{E_i}{\displaystyle \Big[exp\left(\frac{E_i}{T}\right)+\lambda_i\Big]}
\\
-\lambda_i(1-R)\;\frac{E_i}{\displaystyle \Big[exp\left(\frac{E_i}{T}\right)+\lambda_i\Big]^2}\Big],
\end{aligned} 
\end{eqnarray}
Here $y$ is the rapidity variable and $m_T$ is the transverse mass $(m_T=\sqrt{{m}^2+{p_T}^2})$. Also $E_i$ is the energy of ith baryon and $V$ is the total volume of the fireball formed at chemical freeze-out and $N_i$ is the total number of ith baryons. We assume that the freeze-out volume of the fireball for all types of hadrons at the time of the homogeneous emissions of hadrons remains the same. 

By inserting $E_i=m_T\;coshy$ in Eq. (5) and integrating the whole expression over transverse component we can get the rapidity distributions of baryons as follows: 

\begin{equation}
\begin{aligned} 
\Big(\frac{dN_i}{dy}\Big)_{th}=\frac{g_iV\lambda_i}{(2{\pi}^2)}\;\Big[\Big((1-R)-\lambda_i\frac{\partial{R}}{\partial{\lambda_i}}\Big)\; \int_0^\infty \frac{m_T^2\;coshy\;dm_T}{\displaystyle\Big[exp\left(\frac{m_T\;coshy}{T}\right)+\lambda_i\Big]}
\\
-\lambda_i(1-R)\;\int_0^\infty \frac{m_T^2\;coshy\;dm_T}{\displaystyle\Big[exp\left(\frac{m_T\;coshy}{T}\right)+\lambda_i\Big]^2}\Big].
\end{aligned} 
\end{equation}

 Eq.(6) gives the rapidity distributions of baryons arising due to a stationary thermal source. It can be mentioned here that in the above equation, there occurs no free parameter because all the quantities $g$, $V$, $\lambda$, $R$ etc. are determined in the model. Similarly, the rapidity density of mesons can be obtained by using the following formula :

\begin{equation}
\Big(\frac{dN_m}{dy}\Big)_{th}=\frac{g_mV\lambda_m}{(2{\pi}^2)}\;\int_0^\infty \frac{m_T^2\;coshy\;dm_T}{\displaystyle\Big[exp\left(\frac{m_T\;coshy}{T}\right)-\lambda_m\Big]}.
\end{equation}
Here $g_m$, $\lambda_m$ are the degeneracy and fugacity of the meson $m$, respectively and $V$ is the total volume of the fireball at freeze-out.

\subsection{Transverse Mass Spectra}

We use Boltzmann statistics in deriving formula for the transverse mass spectra because we want to calculate spectra of hadrons only at RHIC and LHC energies where the effect of quantum statistics is found to be negligible \cite{Schnedermann:1993}. In the Boltzmann's approximation, Eq.(5) can be reduced to a simple form :

\begin{eqnarray}
\frac{dN_i}{dy\;m_T\;dm_T\;d\phi}=\frac{g_iV\lambda_i}{(2\pi)^3}\;\Big[\Big((1-R)-\lambda_i\frac{\partial{R}}{\partial{\lambda_i}}\Big)\Big] E_i\;\Big[exp\left(\frac{-E_i}{T}\right)\Big].
\end{eqnarray}

Putting $E_i=m_T\;coshy$ in Eq.(8) and integrating over rapidity ($y$), we get the transverse mass spectra as follows :

\begin{eqnarray}
\frac{dN_i}{m_T\;dm_T}=\frac{g_iV\lambda_i}{(2\pi)^3}\;\Big[(1-R)-\lambda_i\frac{\partial{R}}{\partial{\lambda_i}}\Big]
\int_0^\infty m_T\;coshy\;\Big[exp\left(\frac{-m_T\;coshy}{T}\right)\Big]dyd\phi,
\end{eqnarray}

or, :

\begin{eqnarray}
\frac{dN_i}{m_T\;dm_T}=\frac{g_iV\lambda_i}{(2{\pi}^2)}\;\Big[(1-R)-\lambda_i\frac{\partial{R}}{\partial{\lambda_i}}\Big]\;m_T\:K_1\Big(\frac{m_T}{T}\Big),
\end{eqnarray}

where $K_1\displaystyle\Big(\frac{m_T}{T}\Big)$ is the modified Bessel's function :

\begin{eqnarray}
K_1\Big(\frac{m_T}{T}\Big)=\int_0^\infty coshy\;\Big[exp\left(\frac{-m_T\;coshy}{T}\right)\Big]dy.
\end{eqnarray}

Similarly mesonic transverse mass spectra can be evaluated as follows :

\begin{eqnarray}
\frac{dN_m}{m_T\;dm_T}=\frac{g_mV\lambda_m}{(2{\pi}^2)}\;m_T\:K_1\Big(\frac{m_T}{T}\Big).
\end{eqnarray}

\section{Hadronic Spectra with the effect of flow}

In the previous section, we have obtained the expression for rapidity as well as transverse mass spectra arising from a stationary thermal source alone. In this section, we modify the expression for rapidity spectra i.e. Eq. (6), by incorporating a flow velocity in the longitudinal direction. The resulting rapidity spectra of ith hadron, after incorporation of the flow velocity in the longitudinal direction is \cite{Schnedermann:1993,Braun:1996}: 

\begin{eqnarray}
\frac{dN_i}{dy}=\int_{-\eta_{max.}}^{\eta_{max.}} \Big(\frac{dN_i}{dy}\Big)_{th}(y-\eta)\;d\eta,
\end{eqnarray}
where $\displaystyle\Big(\frac{dN_i}{dy}\Big)_{th}$ can be calculated by using Eq.(6) for the baryons and Eq.(7) for the mesons. The average longitudinal velocity used is \cite{Feng:2011,Netrakanti:2005}:
 
\begin{eqnarray}
\langle\beta_L\rangle=tanh\Big(\frac{\eta_{max}}{2}\Big).
\end{eqnarray}
Here $\eta_{max}$ is an important parameter which provides the upper rapidity limit for the longitudinal flow velocity at a particular $\sqrt{s_{NN}}$ and it's value is determined by the best experimental fit. The value of $\eta_{max}$ increases with the increasing $\sqrt{s_{NN}}$ and hence $\beta_L$ also increases.

In the case of transverse mass spectra, we incorporate flow velocity in both the directions, longitudinal as well as transverse, in order to describe the experimental data satisfactorily. However, we take a radial type of flow velocity in the transverse direction which imparts a radial boost on top of the thermal distribution. We thus define the four velocity field in both the directions as follows \cite{Ruuskanen:1987} :  

\begin{eqnarray}
u^{\mu}(\rho,\eta)=(cosh\rho\;cosh\eta,\bar{e_r}\;sinh\rho,cosh\rho\;sinh\eta),
\end{eqnarray}
where $\rho$ is the radial flow velocity in the transverse direction and $\eta$ is the flow velocity in the longitudinal direction. Once we have defined the flow velocity field, we can calculate the invariant momentum spectrum by using the following formula \cite{Schnedermann:1993,Cooper:1974} :

\begin{eqnarray}
E_i\;\frac{d^3N_i}{dk^3}=\frac{g_iV\lambda_i}{(2\pi)^3}\;\Big[(1-R)-\lambda_i\frac{\partial{R}}{\partial{\lambda_i}}\Big]\int exp\Big(\frac{-k_{\mu}u^{\mu}}{T}\Big)\;k_\lambda\;d\sigma_\lambda.
\end{eqnarray}
In the derivation of Eq.(16), we assume that an isotropic thermal distribution of hadrons is boosted by the local fluid velocity $u^{\mu}$. Now the resulting spectrum can be written as \cite{Schnedermann:1993} :

\begin{eqnarray}
\begin{split}
\frac{dN_i}{m_T\;dm_T\;dy}=\frac{g_iV\lambda_i\;m_{T}}{(2\pi)^3}\;\Big[(1-R)-\lambda_i\frac{\partial{R}}{\partial{\lambda_i}}\Big]\;\int r\;dr\;d\phi\;d\zeta
\\
\times\;exp\Big(-\frac{m_T cosh(y-\eta)\;cosh\rho-p_Tsinh\rho\;cos\phi}{T}\Big).
\end{split}
\end{eqnarray}
The freeze-out hypersurface $d\sigma_{\lambda}$ in Eq.(16) is parametrized in cylindrical coordinates $(r,\phi,\zeta)$, where the radius $r$ can lie between $0$ and $R_0$ i.e. the radius of the fireball at freeze-out, the azimuthal angle $\phi$ lies between $0$ and $2\pi$, and the longitudinal space-time rapidity variable $\zeta$ varies between $-\eta_{max}$ and $\eta_{max}$. Now, integrating Eq.(17) over $\phi$ as well as $\zeta$, we get the final expression for the transverse mass spectra \cite{Schnedermann:1993} :

\begin{eqnarray}
\frac{dN_i}{m_Tdm_T}=\frac{g_iV\lambda_i\;m_T}{(2{\pi}^2)}\;\Big[(1-R)-\lambda_i\frac{\partial{R}}{\partial{\lambda_i}}\Big]\int_0^{R_{0}} r\;dr\;K_1\Big(\frac{m_T\;cosh\rho}{T}\Big)I_0\Big(\frac{p_T\;sinh\rho}{T}\Big).
\end{eqnarray}

Here $I_0\displaystyle\Big(\frac{p_T\;sinh\rho}{T}\Big)$ is the modified Bessel's function :

\begin{eqnarray}
I_0\Big(\frac{p_T\;sinh\rho}{T}\Big)=\frac{1}{2\pi}\int_0^{2\pi} exp\Big(\frac{p_T\;sinh\rho\;cos\phi}{T}\Big)d\phi,
\end{eqnarray}
where $\rho$ is given by $\rho=tanh^{-1}\beta_r$, with the velocity profile chosen as $\beta_r=\displaystyle\beta_s\;\Big(\xi\Big)^n$ \cite{Schnedermann:1993,Braun:1996}. $\beta_s$ is the maximum surface velocity and is treated as a free parameter and $\xi=\displaystyle\Big(r/R_0\Big)$. The average of the transverse velocity can be evaluated as \cite{Adcox:2004} :

\begin{eqnarray}
<\beta_r> =\frac{\int \beta_s\xi^n\xi\;d\xi}{\int \xi\;d\xi}=\Big(\frac{2}{2+n}\Big)\beta_s.
\end{eqnarray}

In our calculation we use a linear velocity profile, ($n=1$) and $R_0$ is the maximum radius of the expanding source at freeze-out ($0<\xi<1$) \cite{Adcox:2004}. Similarly following equation can be used to calculate transverse mass spectra for mesons :

\begin{eqnarray}
\frac{dN_m}{m_Tdm_T}=\frac{g_mV\lambda_m\;m_T}{(2{\pi}^2)}\;\int_0^{R_0} r\;dr\;K_1\Big(\frac{m_T\;cosh\rho}{T}\Big)I_0\Big(\frac{p_T\;sinh\rho}{T}\Big). 
\end{eqnarray}

\section{Results and Discussions}

\begin{figure}
\includegraphics[height=22em]{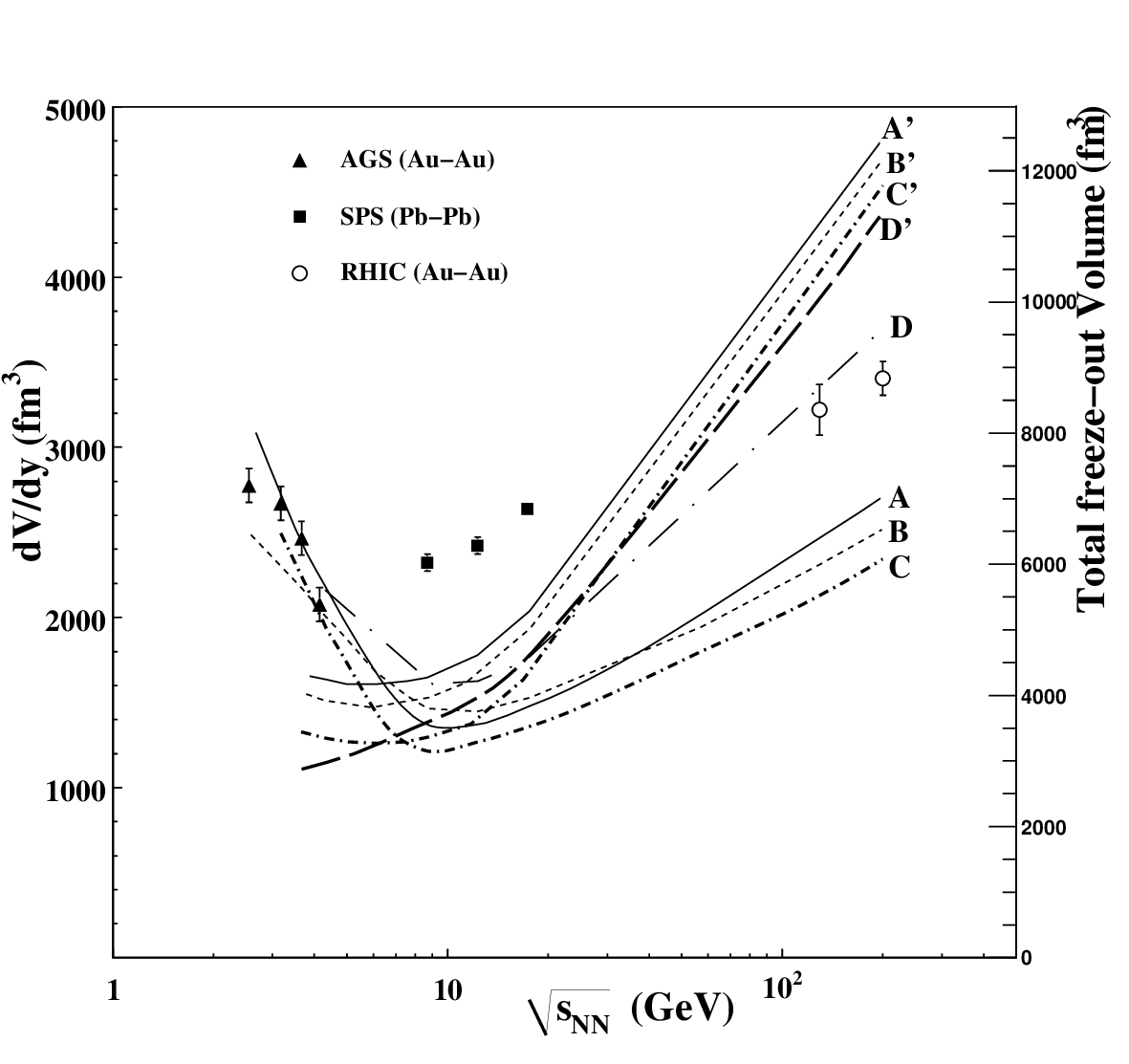}
\caption[]{ Energy dependence of the freeze-out volume for the central nucleus-nucleus collisions. The symbols are the HBT data for freeze-out volume $V_{HBT}$ for the $\pi^+$ \cite{Adamova:2003}. $A'$ ,$B'$ and $C'$ are the total freeze-out volume and $A$ ,$B$ and $C$ depict the $dV/dy$ as found in our model for $\pi^+$, $K^+$ and $K^-$ , respectively. $D$ represents the total freeze-out volume for $\pi^+$ calculated in the Ideal HG model. $D'$ is the the total freeze-out volume for $\pi^+$ in our model calculation using Boltzmann's statistics.}
\end{figure}

\begin{figure}
\includegraphics[height=22em]{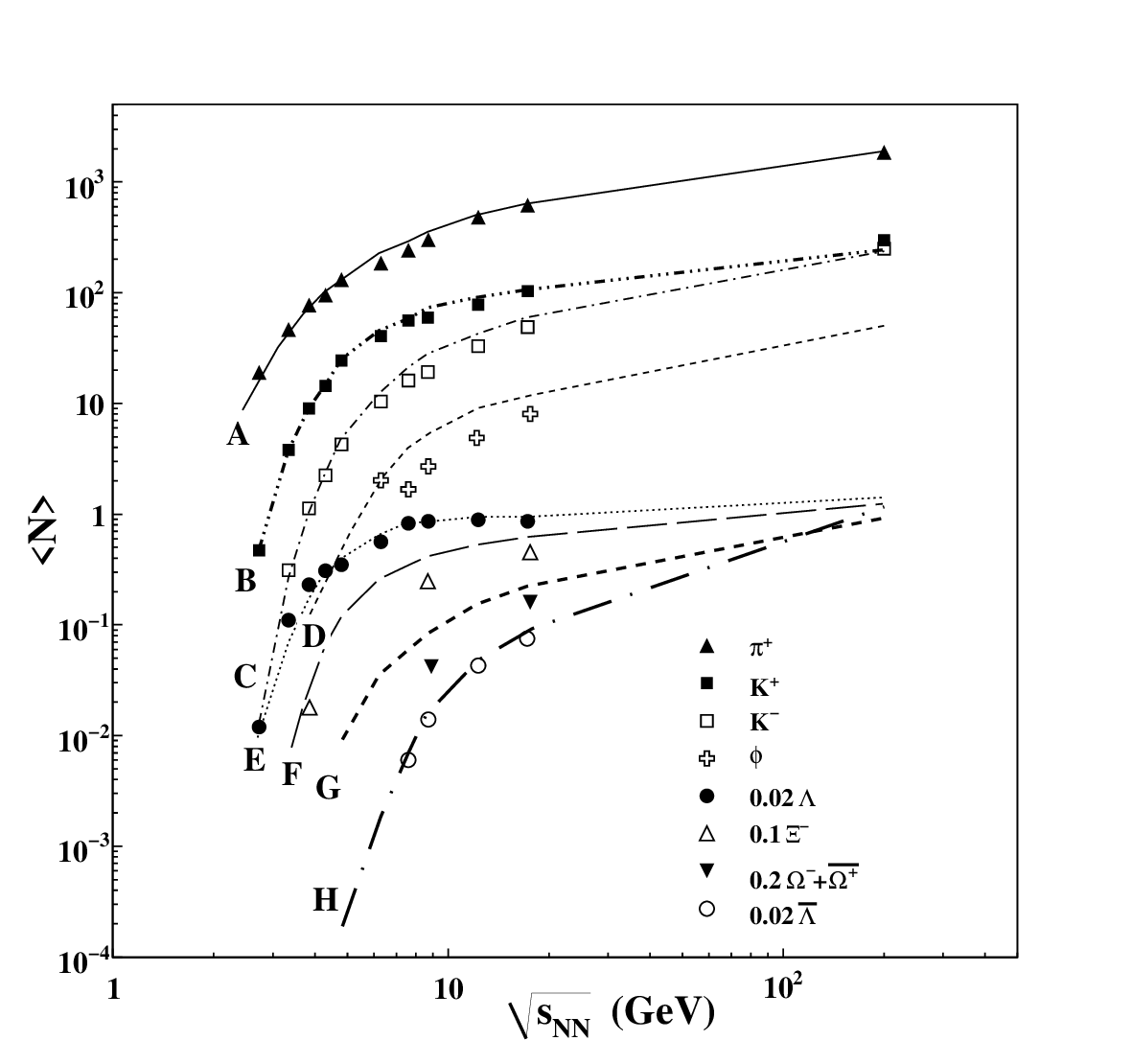}
\caption[]{Variations of total multiplicities of $\pi^+$, $K^+$, $K^-$, $\phi$, $\Lambda$, $\Xi^-$, $(\Omega^{-}+\bar{\Omega^{+}})$, and $\bar{\Lambda}$ with respect to center-of-mass energy predicted by our model. Experimental data measured in central $Au-Au/Pb-Pb$ collisions \cite{Becattini:2001,Klay:2003,Afanasiev:2002,Gazdzicki:2004,Bearden:2005,Anticic:2004,Richard:2005,Afanasiev:2000,Meurer:2004,Afanasiev1:2002,Ahle:1998,Ahle:2000,Ahle1:2000,Ahle1:1998,Pinkenburg:2002,Albergo:2002,Chung:2003} have also been shown for comparison. In this figure, A, B, C, D, E, F, G, and H represent the multiplicities of $\pi^+$, $K^+$, $K^-$, $\phi$, $\Lambda$, $\Xi^-$, $(\Omega^{-}+\bar{\Omega^{+}}$), and $\bar{\Lambda}$, respectively.}
\end{figure}

\begin{figure}
\includegraphics[height=22em]{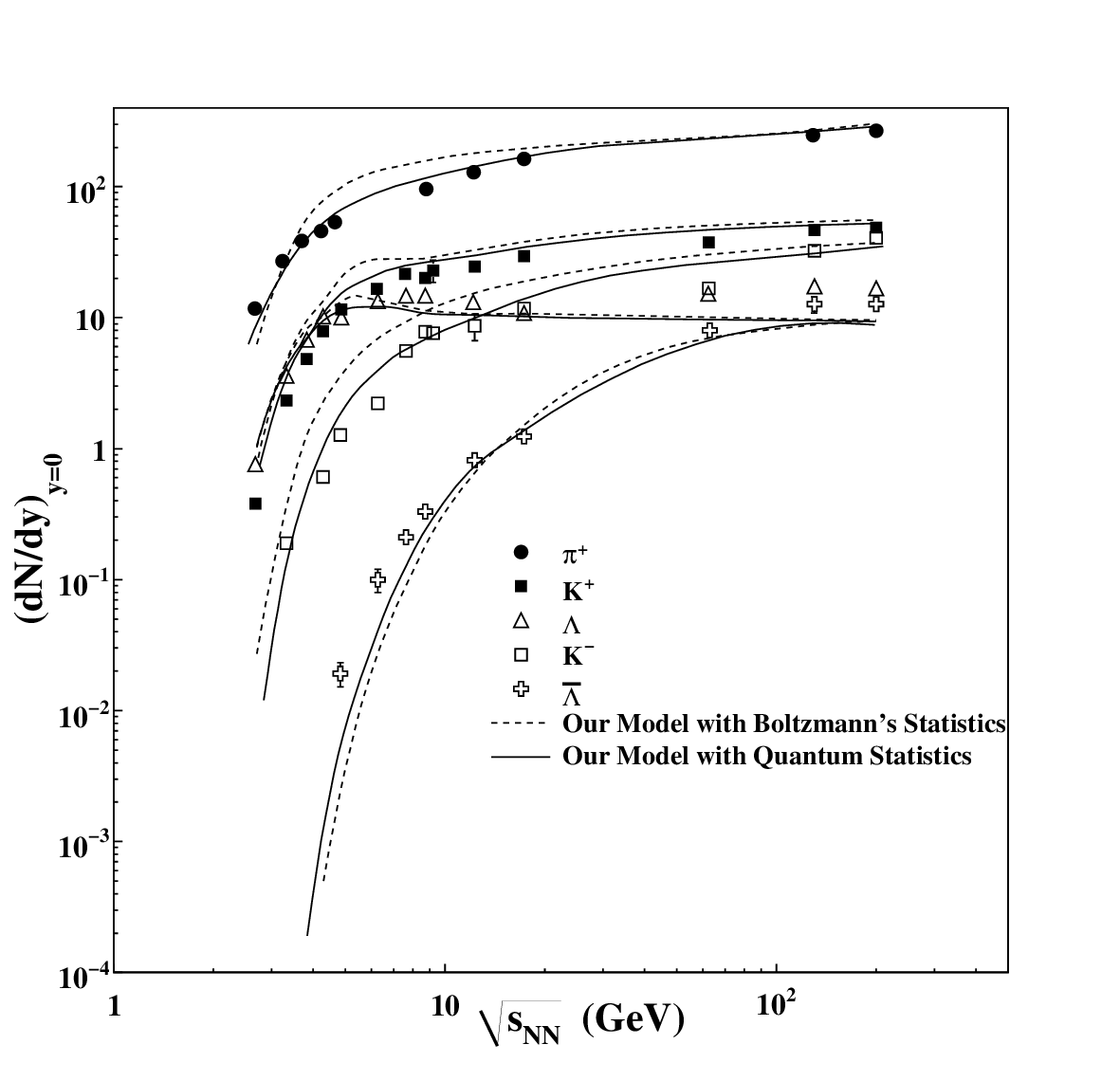}
\caption[]{Variation of rapidity distributions of various hadrons with respect to $\sqrt{s_{NN}}$ at midrapidity. Lines show our model calculation. Symbols are the experimental data \cite{Blume:2011,Andronic:2006,Aggarwal:2011}.}
\end{figure}

We have used freeze-out temperature and baryon chemical potential determined by fitting the particle ratios as described in Ref. \cite{Tiwari:2012}. Then we use a suitable parametrization for $T$ and $\mu_B$ in terms of center-of-mass energies. We have used equal hard-core radius $r'=0.8\;fm$ for all types of baryons. We have considered all the particles and the resonances in the HG upto mass of $2\;GeV/c^2$ in our calculation. We have used resonances having well defined masses, widths etc. and branching ratios for sequential decays are also suitably incorporated.

Although the emission of hadrons from a statistical thermal model essentially invokes the idea of chemical equilibrium, it cannot reveal any information regarding the existence of a QGP phase before hadronization. However, if the constituents of the fireball have gone through a mixed phase, volume $V$ of the fireball at freeze-out is expected to be much larger than what we expect from a system if it remains in the hadronic phase only. In Fig. 1, we have shown $V$ and $dV/dy$ as obtained in our excluded-volume model and their variations with the center-of-mass energy. For this purpose, we have used the data for total multiplicities of $\pi^+$, $K^+$, $K^-$ and after dividing with corresponding number densities obtained in our model, we can get $V$ for $\pi^+$, $K^+$, and $K^-$, respectively. Similarly for deducing $dV/dy$, we use the data for $dN/dy$ and divide respectively by the corresponding number density as calculated in our model. We have compared predictions from our model with the data obtained from the pion interferometry (HBT) \cite{Adamova:2003} which in fact reveals thermal (kinetic) freeze-out volumes. Our results support the finding that the decoupling of strange mesons from the fireball takes place earlier than the $\pi$-mesons. Moreover, a flat minimum occurs in the curves around the center-of-mass energy $\approx8\;GeV$ and this feature is well supported by HBT data. For comparison, we show the total freeze-out volume for $\pi^+$ calculated in our model using Boltzmann's statistics. We see that there is a significant difference between the results arising from quantum statistics and Boltzmann's statistics. We also show the total freeze-out volume for $\pi^+$ in Ideal HG (IHG) model calculation by dash-dotted line $D$. We clearly notice a remarkable difference between the results of our excluded-volume model and that of IHG model also.  

In order to calculate total multiplicities of hadrons, we first determine the total freeze-out volume for $K^+$ by dividing the experimentally measured multiplicities of $K^+$ with it's number density as calculated in our model at different center-of-mass energies. We assume that the fireball after expansion, achieves the stage of chemical equilibrium and the freeze-out volume of the fireball remains same for all particles at the time for their homogeneous emissions. This freeze-out volume thus extracted for $K^+$, has further been used to calculate the multiplicities of all other hadrons from corresponding number densities at different $\sqrt{s_{NN}}$. Figure 2 shows the center-of-mass energy dependence of multiplicities of hadrons $\pi^+$, $K^+$, $K^-$, $\phi$, $\Lambda$, $\Xi^-$, $(\Omega^{-}+\bar{\Omega^{+}})$, and $\bar{\Lambda}$ as predicted by our model calculation. We also show here corresponding experimental data measured in central  $Au-Au/Pb-Pb$ collisions \cite{Becattini:2001,Klay:2003,Afanasiev:2002,Gazdzicki:2004,Bearden:2005,Anticic:2004,Richard:2005,Afanasiev:2000,Meurer:2004,Afanasiev1:2002,Ahle:1998,Ahle:2000,Ahle1:2000,Ahle1:1998,Pinkenburg:2002,Albergo:2002,Chung:2003} for comparison. We observe an excellent agreement between our model calculations and the experimental data for total multiplicities of all particles except $\phi$, $\Xi^-$, and $\Omega^{-}$, where we see small quantitative difference. Again, the thermal multiplicities for all these particles are larger than the experimental values. This analysis thus hints at a new and different mechanism for the production of these strange particles. This conclusion is indeed supported by the results of Linnyk $et\;al.$ \cite{Linnyk:2010}, where the production of multistrange hyperons is shown to be dominated by the contribution from the hadronization of the partonic phase.

\begin{figure}
\includegraphics[height=20em]{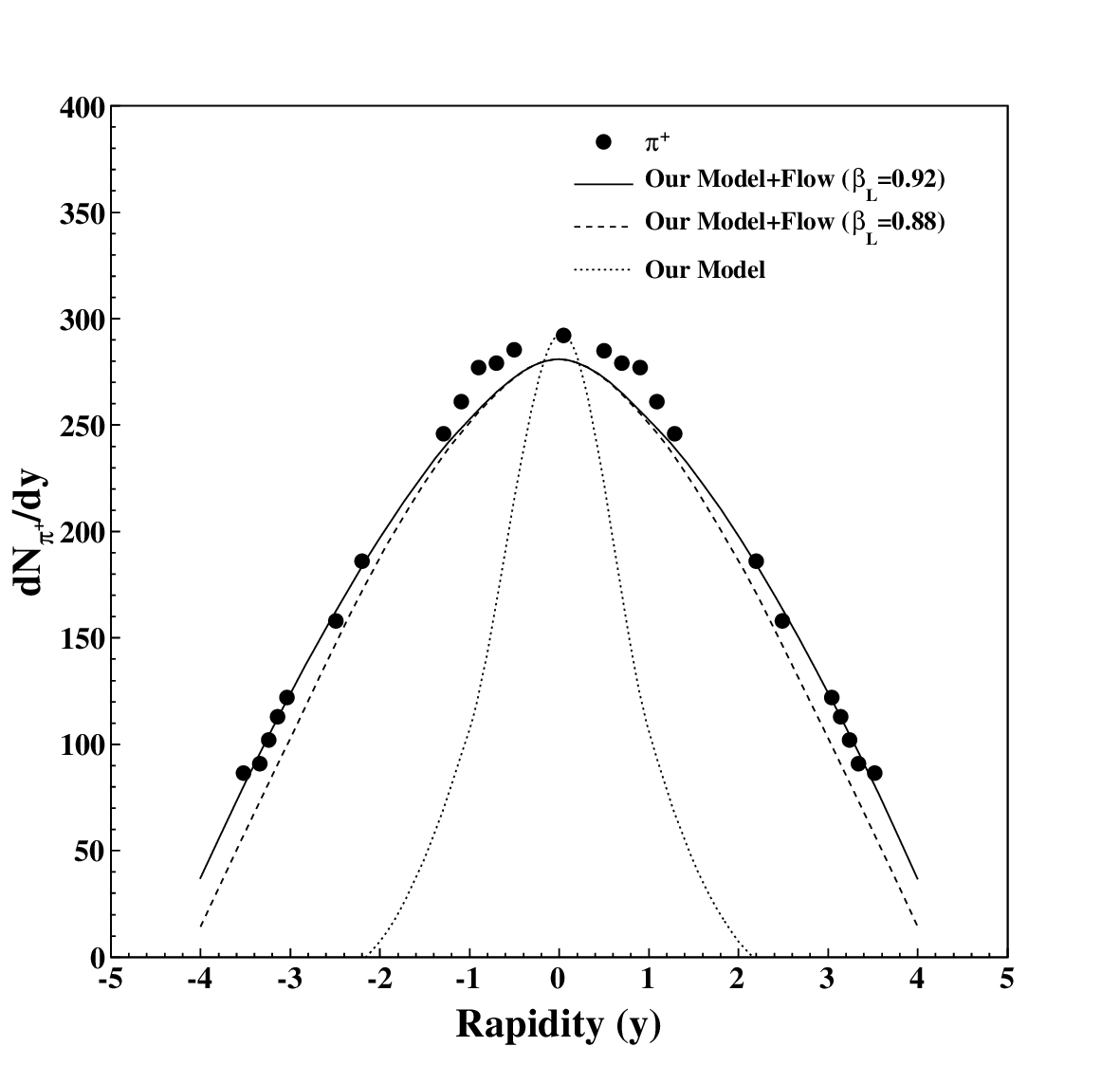}
\caption[]{Rapidity distribution of $\pi^+$ at $\sqrt{s_{NN}}= 200 GeV $. Dotted line shows the rapidity distribution calculated in our thermal model. Solid line and dashed line show the results obtained after incorporating longitudinal flow in our thermal model. Symbols are the experimental data \cite{Bearden1:2005}.}
\end{figure}

In Fig.3, we compare the experimental midrapidity data \cite{Blume:2011,Andronic:2006,Aggarwal:2011} of various hadron species over a broad energy range from AGS to RHIC energies with the results of our model calculations. For each hadron, we use the same freeze-out volume of the fireball as extracted for $K^+$ as mentioned in Fig.2. We also show for comparison the results calculated using Boltzmann's statistics (as done by Mishra $et al.$ \cite{Mishra:2007}). As expected, both results differ only at lower energies. However, the results with full quantum statistics yield a better agreement with the experimental data.

In Fig.4, we present the rapidity distribution of $\pi^+$ for central $Au+Au$ collisions at $\sqrt{s_{NN}}=200\; GeV$ over full rapidity range. Dotted line shows the distribution of $\pi^+$ due to stationary thermal source. Solid line shows the rapidity distributions of $\pi^+$ after the incorporation of longitudinal flow in our thermal model and results give a good agreement with the experimental data \cite{Bearden1:2005}. In fitting the experimental data, we use the value of $\eta_{max}=3.2$ and hence the longitudinal flow velocity $\beta_L=0.92$ at $\sqrt{s_{NN}}=200\; GeV$. For comparison and testing the appropriateness of this parameter, we also show the rapidity distributions at a different value i.e., $\eta_{max}=2.8$ (or, $\beta_L=0.88$), by a dashed line in the figure. We find that the results slightly differ and hence it shows a small dependence on $\eta_{max}$.

\begin{figure}
\includegraphics[height=20em]{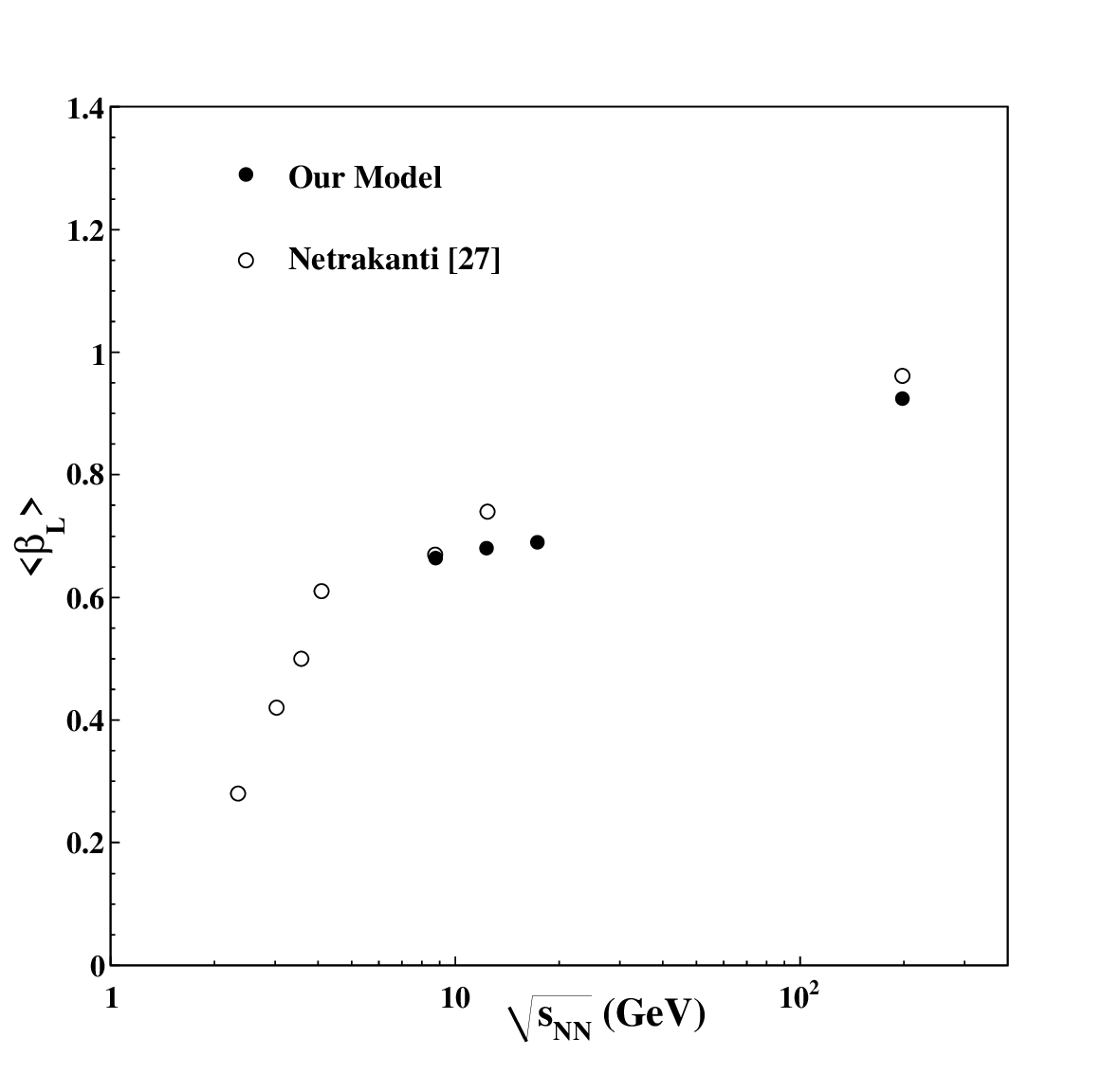} 
\caption[]{Variations of the longitudinal flow velocities with respect to $\sqrt{s_{NN}}$. Open circles are the results of Ref. \cite{Netrakanti:2005}.}
\end{figure}

 Figure 5 demonstrates the variations of longitudinal flow velocities extracted in our model with respect to $\sqrt{s_{NN}}$ and it shows that the longitudinal flow velocity increases with increasing $\sqrt{s_{NN}}$. We compare our results with the values calculated in Ref. \cite{Netrakanti:2005} and find that a small but distinct difference exists between these two results.

\begin{figure}
\includegraphics[height=20em]{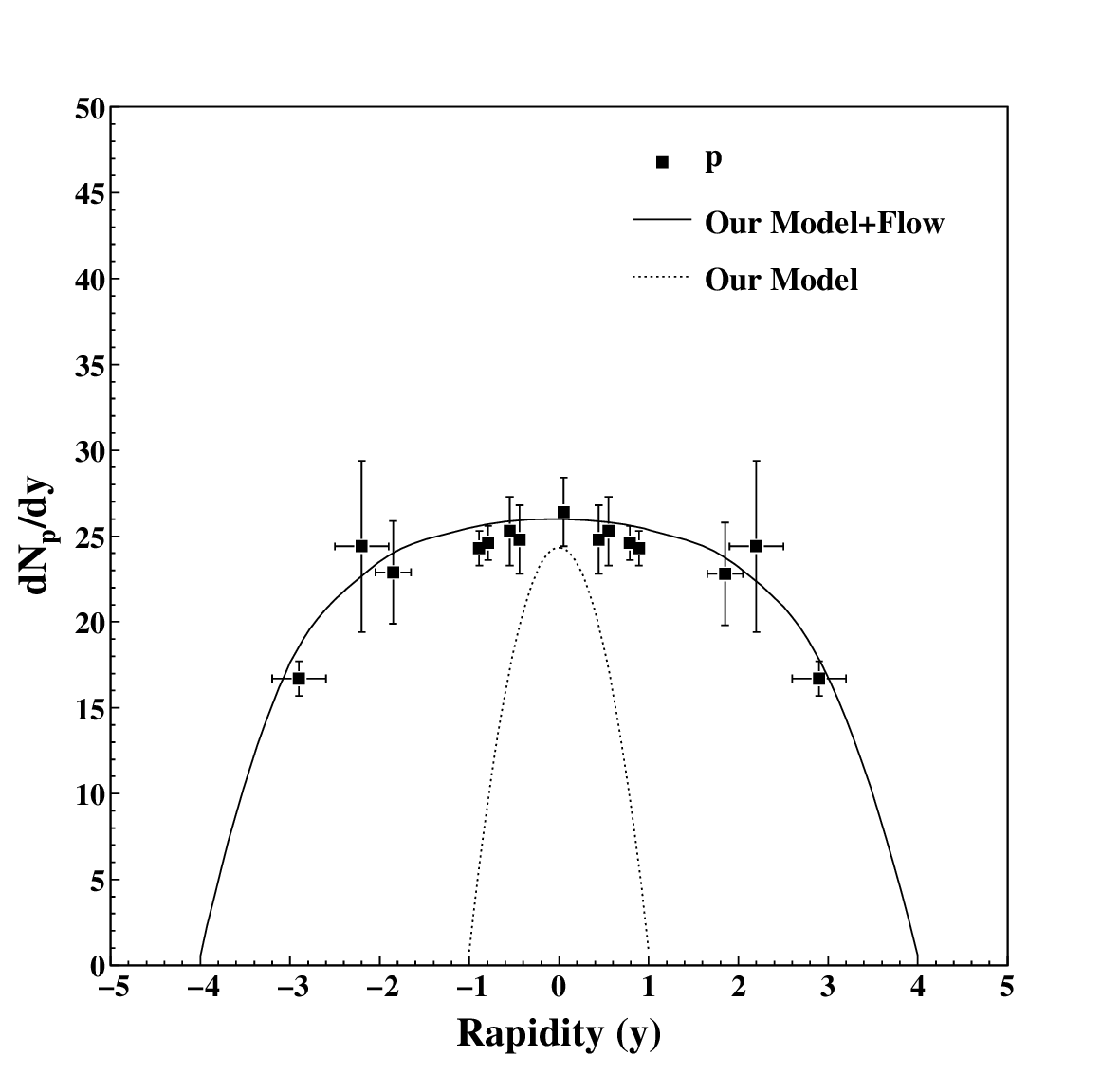}
\caption[]{Rapidity distribution of proton for central $Au-Au$ collisions at $\sqrt{s_{NN}}= 200\; GeV$. Dotted line shows the rapidity distribution due to purely thermal source and solid line shows the result after incorporating the longitudinal flow velocity in our thermal model. Symbols are the experimental data \cite{Bearden:2004}.}
\end{figure}

\begin{figure}
\includegraphics[height=20em]{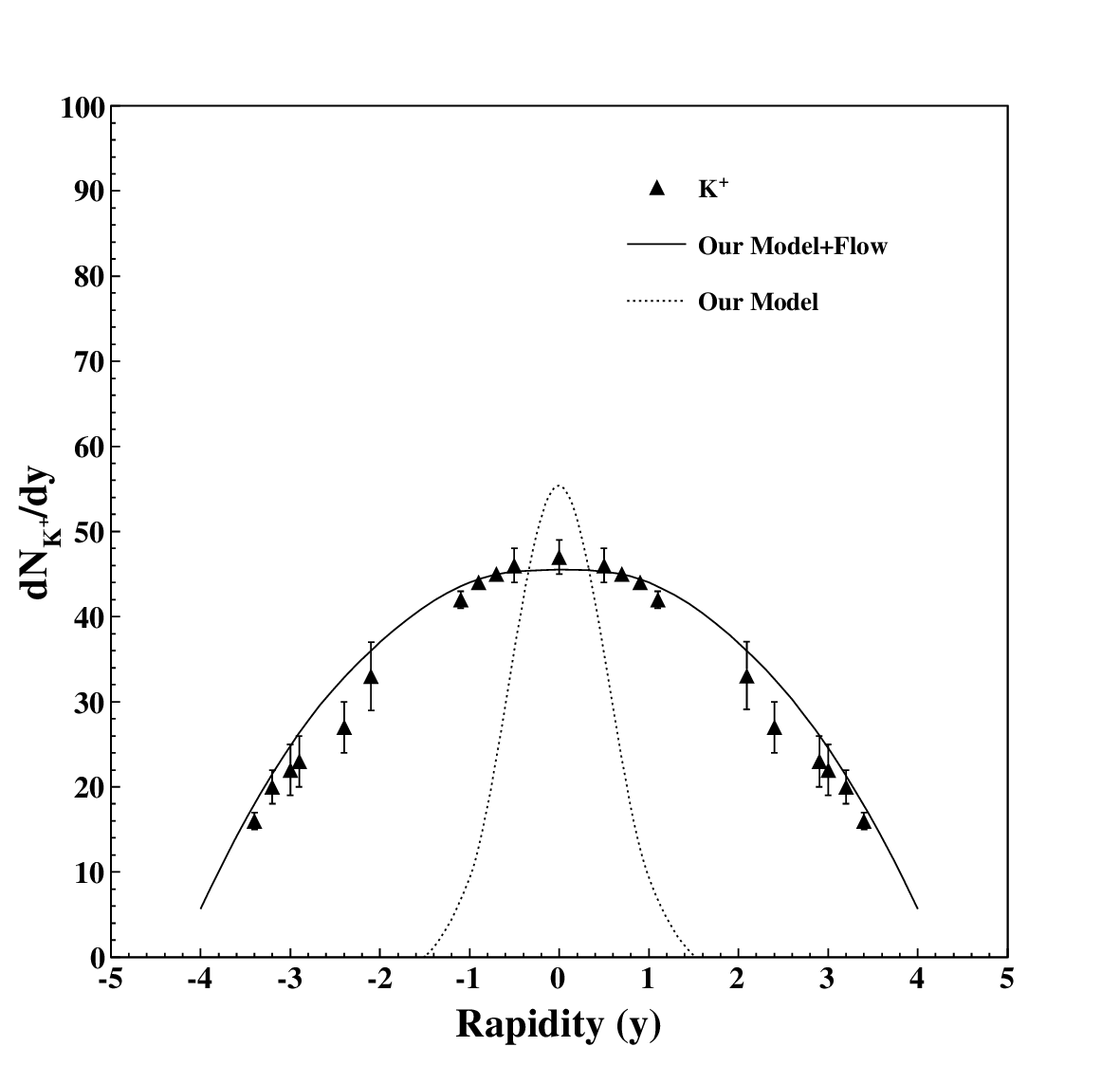}
\caption[]{Rapidity distribution of $K^+$ at $\sqrt{s_{NN}}= 200\;GeV $. Dotted line shows the rapidity distribution due to purely thermal source and solid line shows the result after incorporating longitudinal flow in the thermal model. Symbols are the experimental data \cite{Yang:2008}.}
\end{figure}

Figure 6 demonstrates the rapidity distribution of proton at $\sqrt{s_{NN}}=\;200\;GeV$. Again, our model describes the experimental data \cite{Bearden:2004} successfully. We see that our approach yields a result which matches closely with the experimental result. It should be emphasized that all quantities $eg.$, $V,\;R,\;\displaystyle\frac{\partial{R}}{\partial{\lambda}}$ etc. are precisely calculated in Eq. (6) of our HG model at the time of freeze-out, and hence there is no arbitrary normalizing parameter in the calculation. In contrast, IHG model calculations are normalized with the experimental data in each case separately and hence there is an arbitrariness involved in the quantitative best-fit calculations. Similarly in Figure 7 we show the rapidity distribution for $K^+$  at $\sqrt{s_{NN}}=200\; GeV $. We find a close agreement between our model results and the experimental data \cite{Yang:2008} after we have incorporated the longitudinal flow component in our calculation. We have again used the same value of $\beta_L$ and $V$ for these distributions as well.

\begin{figure}
\includegraphics[height=20em]{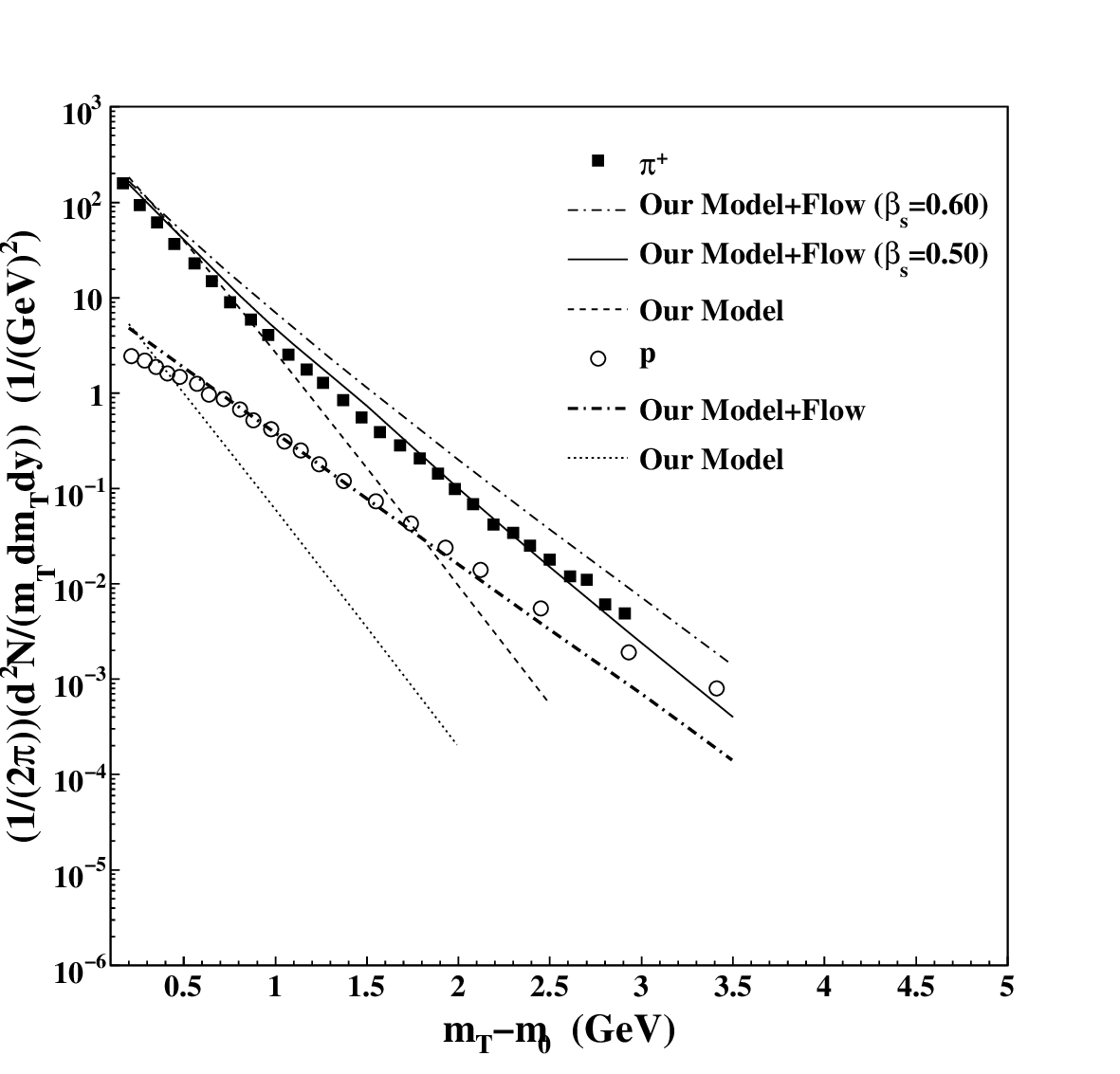}
\caption[]{Transverse mass spectra for $\pi^+$ and proton for the most central collisions at $\sqrt{s_{NN}}= 200\; GeV$. Dashed and dotted lines are the transverse mass spectra due to purely thermal source for $\pi^+$ and proton, respectively. Solid and dash-dotted lines are the results for $\pi^+$ and proton, respectively obtained after incorporation of flow in thermal model. Symbols are the experimental data \cite{Adler:2004}.}
\end{figure}

In Fig. 8, we show the transverse mass spectra for $\pi^+$ and proton for the most central collisions of $Au+Au$ at $\sqrt{s_{NN}}= 200\;GeV$. We have neglected the contributions from the resonance decays in our calculations since these contributions affect the transverse mass spectra only towards the lower transverse mass side i.e. $m_T<0.3 \;GeV$. Our model calculations show some difference with the experimental data for $m_T<0.3 \;GeV$ but a good agreement between our calculations and the experimental results for $m_T>0.3 \;GeV$ is demonstrated after we have incorporated the resulting flow effect. This again shows the importance of collective flow in the description of the experimental data \cite{Adler:2004}. At this energy, the value of $\beta_s$ is taken as $0.50$ and transverse flow velocity $\beta_r=0.33$. This set of transverse flow velocity is able to reproduce the transverse mass spectra of almost all the hadrons at $\sqrt{s_{NN}}= 200\; GeV$. We notice that the transverse flow velocity slowly increases with the increasing $\sqrt{s_{NN}}$. If we take $\beta_s=0.60$, we find that the results differ with data as shown in Fig. 8.

\begin{figure}
\includegraphics[height=20em]{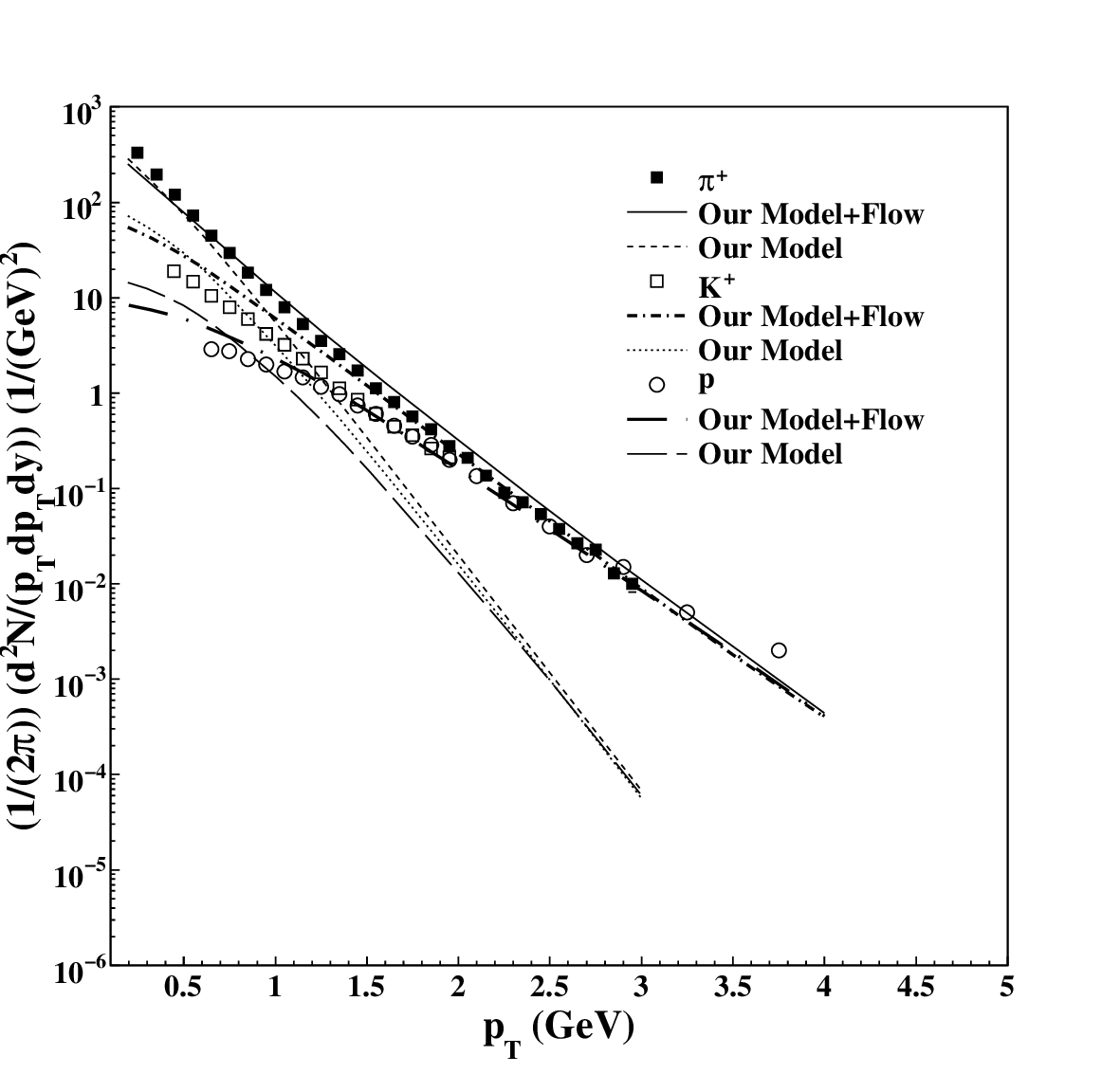}
\caption[]{Transverse momentum spectra for $\pi^+$, $p$, and $K^+$ for the most central $Au-Au$ collision at $\sqrt{s_{NN}}=200\:GeV$. Lines are the results of our model calculation and symbols are the experimental results \cite{Adler:2004}.}
\end{figure}

\begin{figure}
\includegraphics[height=22em]{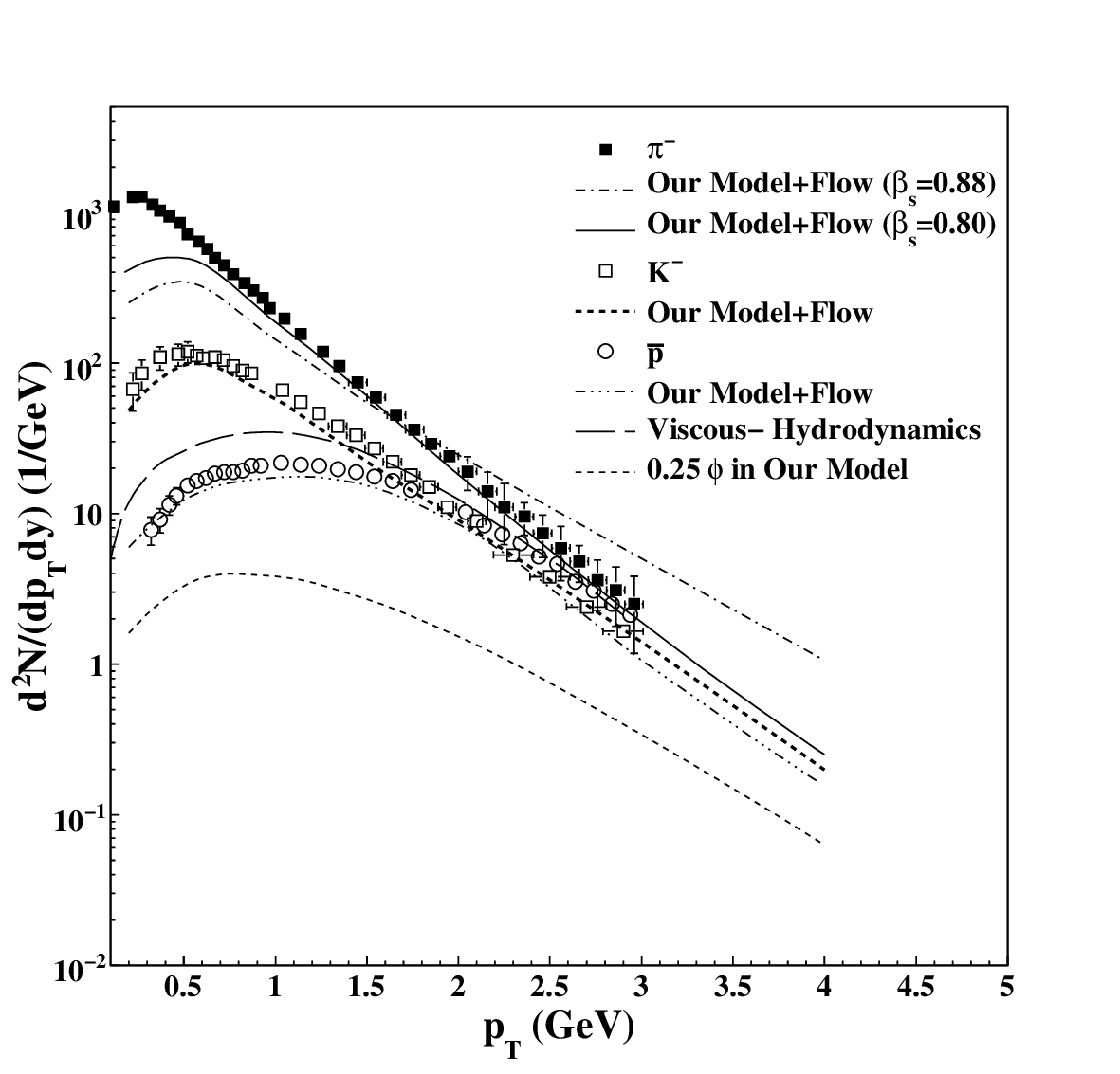}
\caption[]{Transverse momentum spectra of various hadrons for the most central collisions of $Pb-Pb$ at $\sqrt{s_{NN}}=2.76\;TeV$ from LHC. Lines are the results of model calculations and symbols are the experimental results \cite{Floris:2011}. Thick-dashed line is the prediction of viscous-hydrodynamical model \cite{Shen:2011} for $\bar{p}$. Dashed line is the prediction from our calculation for $\phi$ meson. }
\end{figure}

We also present the $p_{T}$- spectra of hadrons at various center-of-mass energies. In Fig. 9, we show the $(p_{T})$ spectra for $\pi^{+}$, $K^+$ and $p$ in the most central collisions of $Au-Au$ at $\sqrt{s_{NN}}=200\;GeV$. Our model calculations reveal a close agreement with the experimental data \cite{Adler:2004}. In Fig. 10, we show the $p_{T}$ spectra of $\pi^{-}$, $K^-$ and $\bar{p}$ for the $Pb-Pb$ collisions at $\sqrt{s_{NN}}=2.76\;TeV$ at the LHC. Our calculations again give a good fit to the experimental results \cite{Floris:2011}. We also compare our results for $\bar{p}$ spectrum with the hydrodynamical model of Shen $et\; al.$ \cite{Shen:2011}, which successfully explains $\pi^{-}$, and $K^-$ spectra but strongly fails in the case of $\bar{p}$ spectrum. In comparison, our model results show closer agreement with the experimental data. Shen $et\; al.$ \cite{Shen:2011} have employed $(2+1)$-dimensional viscous hydrodynamics with the lattice QCD-based EOS. They use Cooper-Frye prescription to implement kinetic freeze-out in converting the hydrodynamic output into the particle spectra. Due to lack of a proper theoretical and phenomenological knowledge, they use the same parameters for $Pb-Pb$ collisions at LHC energy, which was used for $Au-Au$ collisions at $\sqrt{s_{NN}}=200\;GeV$. Furthermore, they use the temperature independent $\eta/s$ ratio in their calculation. After fitting the experimental data, we get $\beta_{s}=0.80$ $(\beta_{r}=0.53)$ at this energy which indicates the collective flow becoming stronger at LHC energy than that observed at RHIC energies. We also predict the $p_{T}$ spectra for $\phi$ meson at this energy. In this plot, we also attempt to show how the spectra for $\pi^-$ will change at a slightly different value of the parameter $i. e.$, $\beta_s=0.88$.

\section{Production of light nuclei, hypernuclei, and antinuclei.}

The hot and dense matter created in ultra-relativistic heavy-ion collisions is uniquely suitable for the study of production of light nuclei, hypernuclei and their antinuclei. Theoretical calculations indicate that the thermal model as well as the coalescence model can describe the relative production abundance of such objects in high energy heavy-ion collisions \cite{Andronic:2011,Andronic1:2011,Cleymans:2011,Xue:2012}. Such an exercise certainly helps in understanding the creation of matter-antimatter asymmetry arising in the early universe and also the strength of nuclear interaction for the antinuclei. It also gives a hint on the degree of thermalization for heavy nuclei in the fireball created after the heavy-ion collisions. Several authors $e.g.$, Andronic $et \; al.$ \cite{Andronic:2011,Andronic1:2011}, Cleymans $et \; al.$ \cite{Cleymans:2011}, and Xue  $et \; al.$ \cite{Xue:2012} have recently attempted to explain the production of nuclei, antinuclei and hypernuclei using thermal and simple coalescence model calculations. It is worthwhile to study the results on the basis of our model and to compare them with those from other calculations.

\begin{figure}
\includegraphics[height=22em]{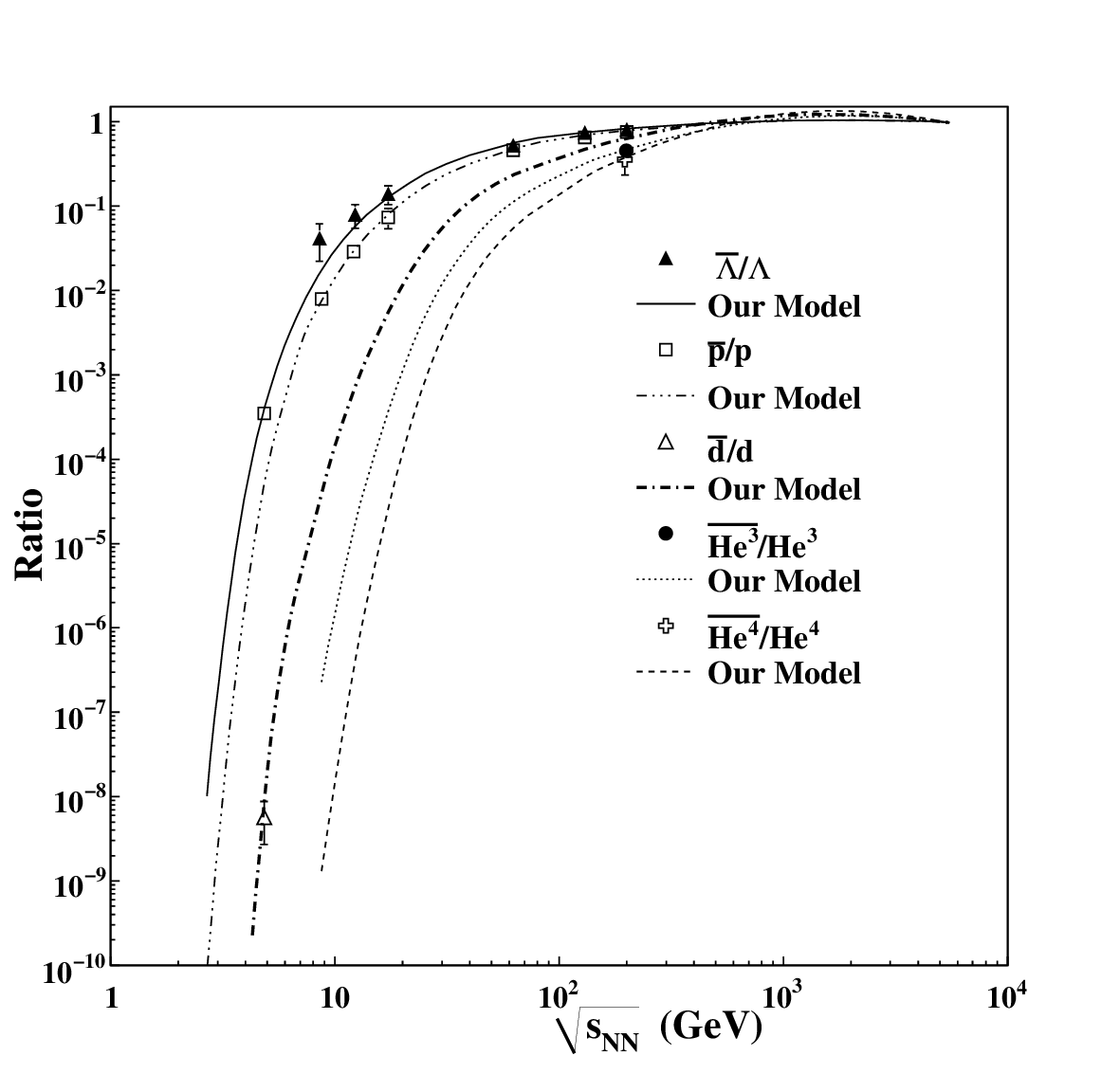}
\caption[]{The energy dependence of anti-baryon to baryon yield and antinuclides to nuclides ratios. Lines are our model calculation and symbols represent the experimental data \cite{Andronic:2011,Andronic1:2011}.}
\end{figure}

\begin{figure}
\includegraphics[height=22em]{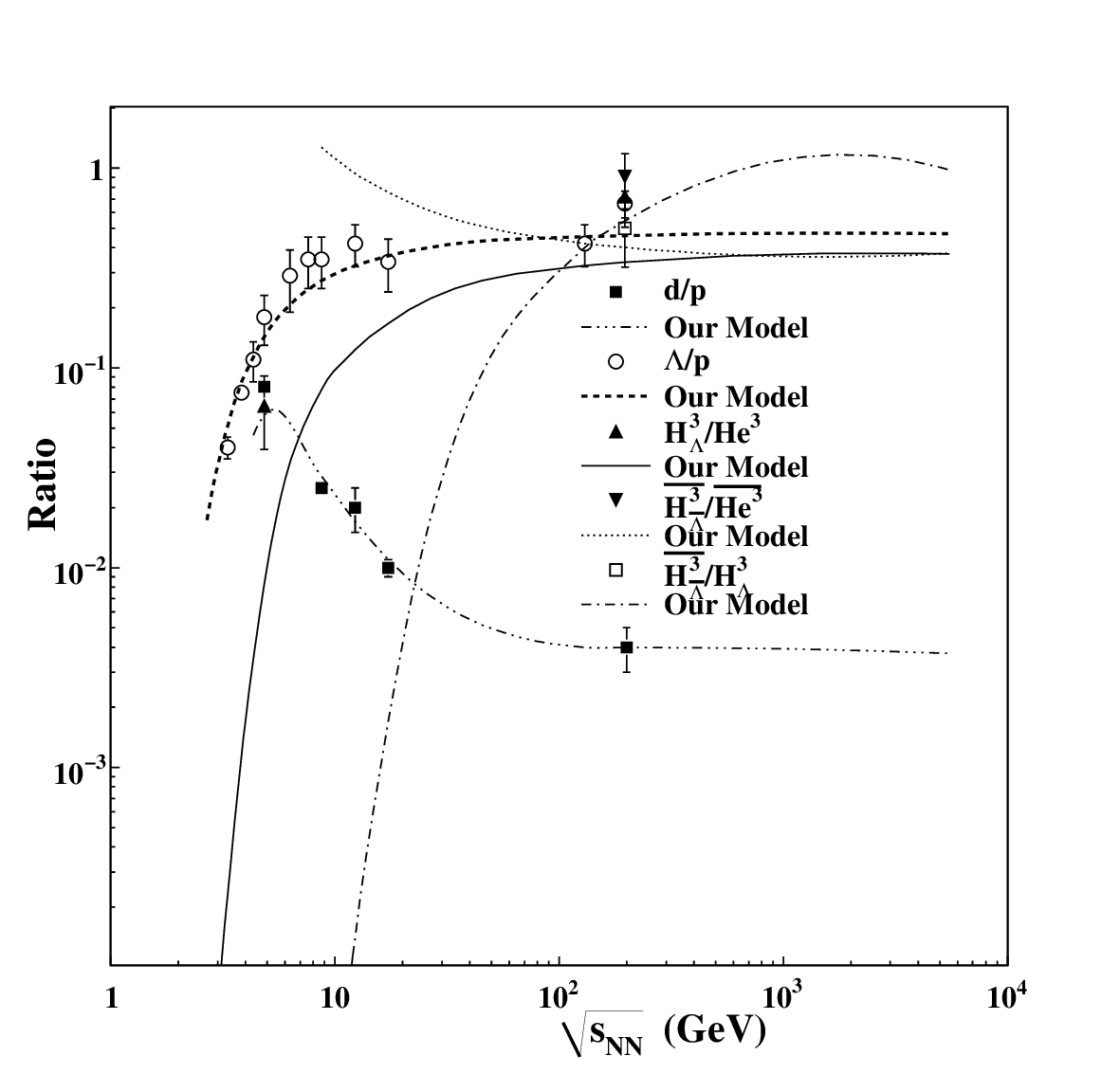}
\caption[]{The energy dependence of various baryons, antibaryons, nuclei, and antinuclei yield ratios. Lines are our model calculation and symbols represent the experimental results \cite{Andronic:2011,Andronic1:2011}.}
\end{figure}

We present an analysis of light nuclei, hypernuclei and their anti-particles using the chemical freeze-out concept within our model calculation to see the effect of excluded-volume picture. The productions of light nuclei and hypernuclei at chemical freeze-out points may not be appropriately calculated, because their binding energies are of the order of few MeV and the chemical freeze-out temperatures are around $100-165 \;MeV$. But we know that the relative yield of particles composed of nucleons is mainly determined by the entropy per baryon which is fixed at chemical freeze-out line in our model \cite{Tiwari:2012}. This was first outlined in \cite{Siemens:1979} and was subsequently emphasized in \cite{Hahn:1988}. This thus constitutes the basis of thermal analyses for the yields of light nuclei \cite{Braun:1995,Braun:2002}. Thus the production yields of light nuclei and hypernuclei is entirely governed by the entropy conservation. Recently the first measurement of the lightest (anti) hypernuclei was done by the STAR experiment at RHIC \cite{Abelev:2010}. We ask an interesting question here whether the production of antinuclei can also be explained by our model without invoking any other parameter. In this section, we have calculated the production yields of light nuclei, hypernuclei, and heavy baryons (anti-baryons) within our thermal model approach and have compared our thermal model calculations with the experimental data. The freeze-out parameters for these yields are taken as the same as was used for other ratios \cite{Tiwari:2012}.

In Fig. 11, we show the energy dependence of $\bar p/p, \bar\Lambda/\Lambda, \bar He^3/He^3, \bar He^4/He^4 $ and $\bar d/d$ yield ratios over a very broad energy range. We compare our results with the experimental data \cite{Andronic:2011,Andronic1:2011} and find a reasonable agreement between these two. However, such ratios of equal mass particles do not involve significant changes when we use excluded-volume model in comparison to IHG calculations. We again stress that the agreement between the experiment and the calculated curves for the ratio $\bar He^3/He^3$ demonstrates the important role of entropy conservation in the analysis of the production of light nuclei as taken in our model. In Fig. 12, we show the energy dependence of the yield ratios $\Lambda/p$, $d/p$ and on comparison, we find that our results reproduce the main features of the experimental data quite well. Similarly we have also shown in Fig. 12 the results of our calculation for the variation of the ratios  $H_{\Lambda}^3/He^3$, $\bar H_{\bar\Lambda}^3/H_{\Lambda}^3$ and $\bar H_{\bar\Lambda}^3/\bar He^3$ with $\sqrt{s_{NN}}$ and we hope in near future, more experimental data will appear to verify our predictions. We thus conclude that the incorporation of excluded-volume effect gives a suitable thermal model calculation for heavier particles, nuclei, hypernuclei, and their antinuclei, also.

\section{Summary and Conclusions}
In summary,  we find that our model provides a good fit to the variations of total multiplicities as well as mid-rapidity densities of various particles and we deduce a large freeze-out volume of the fireball at RHIC energy and this picture supports the idea of a mixed phase after QGP formation before hadronization because a huge size of a homogeneous fireball source can only arise if a mixed phase has occurred before the formation of a hot, dense HG. Further, we present an analysis of rapidity distributions and transverse mass spectra of hadrons in central nucleus-nucleus collisions at various center-of-mass energies using our equation of state (EOS) for HG. We see that the stationary thermal source alone cannot describe the experimental data fully unless we incorporate flow velocities in the longitudinal as well as in the transverse direction and as a result our modified model predictions show a  good agreement with the experimental data. Our analysis shows that a collective flow develops at each $\sqrt{s_{NN}}$ which increases further with the increasing $\sqrt{s_{NN}}$. The description of the rapidity distributions and transverse mass spectra of hadrons at each $\sqrt{s_{NN}}$ matches very well with the experimental data. Although, we are not able to describe successfully the spectra for multistrange particles which suggests that a somewhat different type of mechanism is required in these cases \cite{Linnyk:2010}. We find that the particle yields and ratios measured in heavy-ion collisions are described well by our thermal model and show an overwhelming evidence for a chemical equilibrium at all beam energies. The rapidity distributions and transverse mass spectra which essentially are dependent on thermal parameters, also show a systematic behaviour and their interpretations most clearly involve the presence of a collective flow in the description of the thermal model. 

In conclusion, we have formulated a thermodynamically consistent excluded-volume model in which geometrical hard-core volume of each baryon gives rise to excluded-volume effect. Our model describes successfully thermal and transport quantities \cite{Tiwari:2012} and gives an explanation for the particle ratios, freeze-out volume, total multiplicities, midrapidity density of the particles and rapidity as well as transverse mass spectra of various hadrons. Recently Gorenstein criticized our paper on the front of thermodynamical consistency \cite{Gorenstein:2012}. We have obtained excluded-volume correction by performing explicit integration over the ``available'' volume in the grand canonical partition function and we have attempted to derive all thermodynamical quantities directly from the partition function \cite{Tiwari:2012} which guarantees thermodynamical consistency. However, we have used an approximation in the form of Neumann iteration and retained only the lowest order terms in deriving quantities. A slight discrepancy in the thermodynamical consistency of the quantities may arise due to this reason. In a future work, we will examine this question in detail \cite{Tiwari:2013}.

\section{ACKNOWLEDGMENTS}

S.K.T. and P.K.S. are grateful to the Council of Scientific and Industrial Research (CSIR), New Delhi, and University Grants Commission (UGC), New Delhi for providing research grants.

\end{document}